# Millimeter-Wave Communication Testbed Using Digital Coding Dynamic Metasurface Antenna: Practical Design and Implementation

Abdul Jabbar*, Mostafa Elsayed, Jalil Ur-Rehman Kazim, Zhibo Pang, Julien Le Kernec, Muhammad Imran, Hadi Larijani, Masood Ur-Rehman, Qammer Abbasi*, and Muhammad Usman

***Abstract*—Dynamic Metasurface Antennas (DMAs) are transforming reconfigurable antenna technology by enabling energy-efficient, cost-effective beamforming through programmable meta-elements, eliminating the need for traditional phase shifters and delay lines. This breakthrough technology is emerging to revolutionize beamforming for next-generation wireless communication and sensing networks. In this paper, we present the design and real-world implementation of a DMA-assisted wireless communication platform operating in the license-free 60 GHz mmWave band. Our system employs high-speed binary-coded sequences generated via field programmable gate array (FPGA), enabling real-time beam-steering for spatial multiplexing and independent data transmission. A proof-of-concept experiment successfully demonstrates high-definition QPSK modulated video transmission at 62 GHz. Moreover, leveraging the DMA's multi-beam capability, we simultaneously transmit video to two spatially separated receivers, achieving accurate demodulation. We envision the proposed mmWave testbed as a platform for enabling the seamless integration of sensing and communication by allowing video transmission to be replaced with sensing data or utilizing an auxiliary wireless channel to transmit sensing information to multiple receivers. This synergy paves the way for advancing integrated sensing and communication (ISAC) in beyond 5G and 6G networks. Moreover, our testbed demonstrates its potential for future real-world use cases, including mmWave backhaul links and massive MIMO mmWave base stations.**

Abdul Jabbar, Hadi Larijani, and Muhammad Usman are with the School of Computing, Engineering and Built Environment, Glasgow Caledonian University, UK (email: abduljabbar@ieee.org; h.larijani@gcu.ac.uk; muhammad.usman@gcu.ac.uk

M. Elsayed, Jalil Ur-Rehman Kazim, Julien Le Kernec, Muhammad Imran, and Masood Ur-Rehman are with James Watt School of Engineering, University of Glasgow, UK (email: mostafa.elsayed; jalil.kazim; julien.lekernec; muhammad.imran; masood.urrehman @glasgow.ac.uk)

Qammer Abbasi is with the University of Glasgow, UK, and also as Distinguished Visiting Research Fellow Abu Dhabi University's College of Engineering (*corresponding author; email: qammer.abbasi@glasgow.ac.uk)

Zhibo Pang is with ABB Corporate Research Center, Vasteras, Sweden, and also with KTH Sweden (email: zhibo@kth.se)

(*corresponding authors: qammer.abbasi@glasgow.ac.uk, abduljabbar@ieee.org)

## I. Introduction

DYNAMIC metasurface antennas (DMA) are envisioned as a groundbreaking class of next-generation reconfigurable antennas offering remarkable beamforming and beam-steering capabilities through their innovative architecture and performance. Unlike conventional phased array antenna systems and active electronically scanned array antennas that rely on complex feed networks and phase shifters, DMAs eliminate these components and leverage software-controlled programmability of individual meta-elements for real-time beam-steering capabilities. The readily available dynamic control of their constituent meta-elements helps to achieve adaptive radiation control and manipulate beams in a variety of ways [1]–[3]. DMAs have shown the potential to enable efficient and flexible beamforming while consuming approximately 30% less power compared to traditional phased array antenna systems [3], [4]. This versatile radiative capability of the metasurface aperture through software control is the hallmark of DMAs, making them a highly promising and enabling candidate for cutting-edge future sixth-generation (6G) wireless networks [5].

Metasurfaces are realized as two-dimensional artificial structures composed of precisely engineered subwavelength elements, which exhibit remarkable capabilities in controlling the amplitude, phase, and polarization of electromagnetic (EM) waves across both transmission and reflection modes [6]–[8]. Beyond their diverse physical phenomena such as invisible cloaking, polarization control, and imaging; programmable metasurfaces are highly anticipated to play a crucial role in next-generation wireless technologies. Such metasurfaces, based on the topological performance, are referred in the literature as programmable coded metasurfaces [9], [10], reconfigurable intelligent surfaces (RIS) [11], [12], and beyond-diagonal RIS [13]. Very recently, programmable metasurfaces have demonstrated striking wave phenomena that are unlocking transformative capabilities in next-generation wireless systems. These advancements encompass intelligent and physically secure wireless communication with direct modulation techniques [14]–[20], non-invasive sensing and monitoring [21], cognitive EM agents for dynamic wave control [22], [23], novel radar architecture [24], as well as integrated sensing and communication (ISAC) [25]–[27].

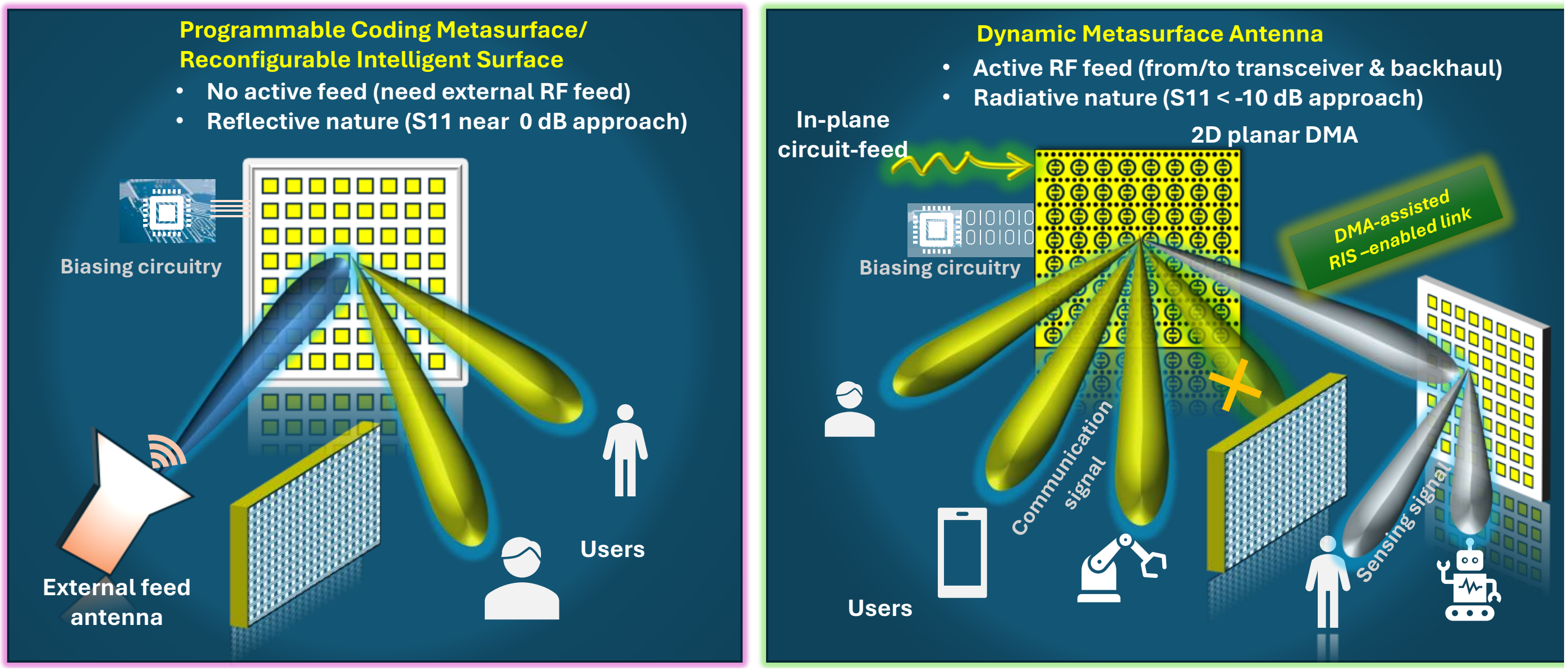

Fig. 1. (left) Application scenario of *reflective* coded metasurfaces (such as RIS) for signal redirection. (right) Conceptual depiction of a planar 2D radiative DMA architecture having an in-plane feed. Moreover, DMA is exhibited to provide a dedicated feed signal towards RIS for NLoS cases. This enables to devise a DMA-assisted RIS-enabled robust wireless communication and sensing paradigm.

Programmable metasurfaces are often reflective in nature, though certain design considerations also enable transmissive configurations [28], [29]. Reflection and transmission-type metasurfaces are typically fed by external antennas, such as a horn antenna, which introduces several limitations such as obstruction of the reflected wave and reducing efficiency. Additionally, both types require a certain free-space distance for proper illumination, leading to bulky and high-profile designs. These constraints make traditional reflective and transmissive metasurface-based systems unreasonable for compact and lightweight applications.

Unlike externally illuminated reflective metasurfaces, DMAs are radiative in nature and circuit-fed from the side of the waveguide aperture in a series-fed manner to generate (and absorb) EM waves [4], [5], [30]–[33]. The EM wave propagates along the metasurface aperture and excites the DMA meta-elements one by one at their designed operational frequency band, leading to the phenomenon of holographic beamforming [34]. Hence, large DMA arrays result in a significantly simpler wiring layout using standard printed circuit board (PCB) technology compared to traditional corporate-fed phased arrays [35]. The traditional corporate-fed antenna systems require individual feeding lines for each antenna element in a parallel configuration which are complex and often lossy in nature.

### *A. A brief Outlook of State-of-the-Art*

In light of the increasingly rigorous demands of 6G wireless networks where array gain, diversity gain, and interference mitigation are essential for enhancing capacity and spectral efficiency, DMAs offer the potential to integrate a large number of radiating elements within a compact surface area [1]. This leads to their enabling role in the evolution of gigantic/ultra-massive multiple-input multiple-output (mMIMO) technology for intelligent base stations [35]–[37]. More specifically, at high-frequency regime where path loss is high and antenna size is reduced, large antenna arrays are highly desired where DMA-based base stations have the potential to employ a large array of antennas with a simplified feed structure and less expensive hybrid beamforming. Consequently, systems can achieve significant enhancements toward spectral efficiency and spatial beamforming gain while maintaining relatively simple signal processing [33], [36], [38].

It is important to emphasize that DMAs can complement reflective coded metasurfaces/RIS, particularly at high-frequency bands, by enhancing RIS functionality and enabling seamless wireless connectivity. For instance, a DMA can serve as an intelligent excitation antenna source with predefined beamforming and coding information that could steer the beam dedicated to the RIS location. Consequently, in non-line-of-sight (NLoS) scenarios, a reliable wireless communication link can be guaranteed as the DMA dynamically steers its beam toward the RIS in real-time. This synergy holds the potential to enable the development of a robust future communication framework, which can be envisioned as a *DMA-assisted RIS-enabled* wireless communication paradigm, as illustrated in Fig. 1.

Furthermore, ISAC can be effectively realized using DMA, where a DMA-assisted link for RIS can facilitate sensing around the corners and blind spots, while the other unobstructive DMA beams can be strategically utilized for direct LoS communication, enabling seamless integration of sensing and communication within the same framework.

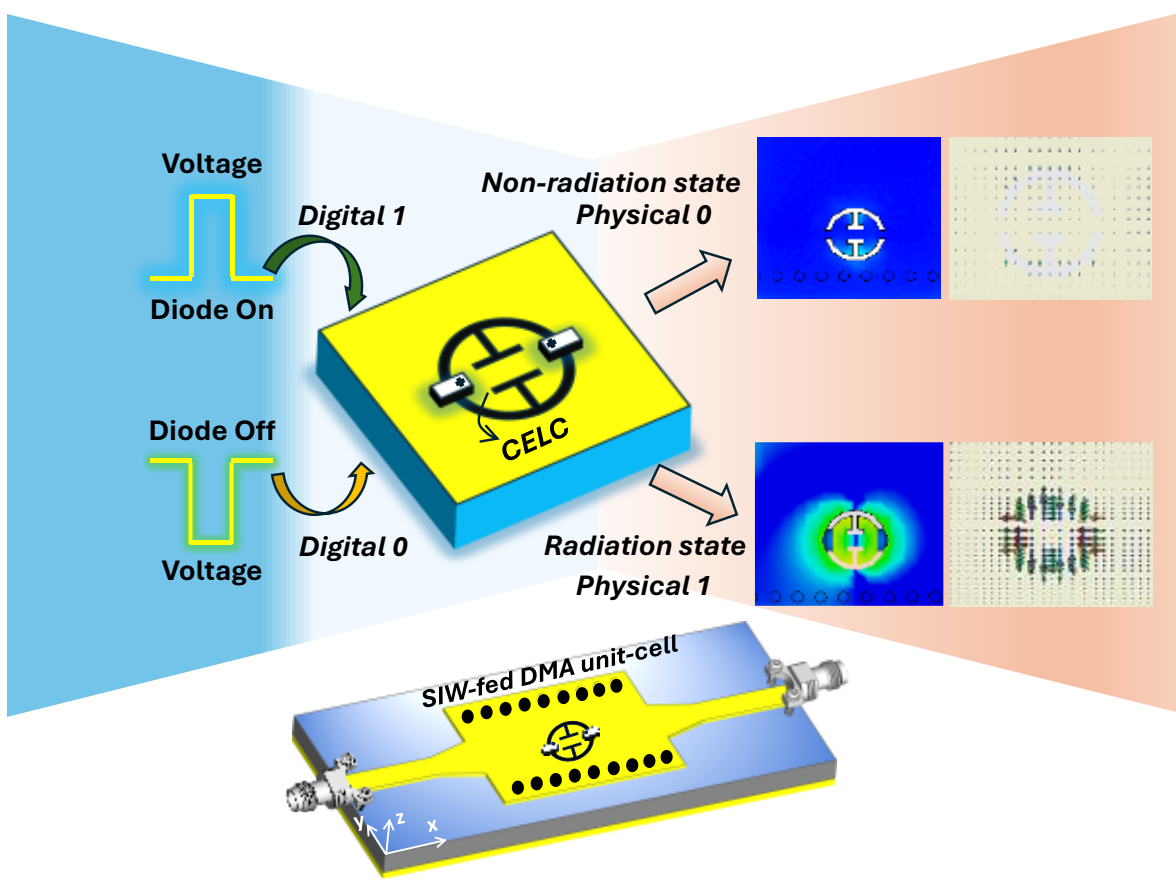

(a)

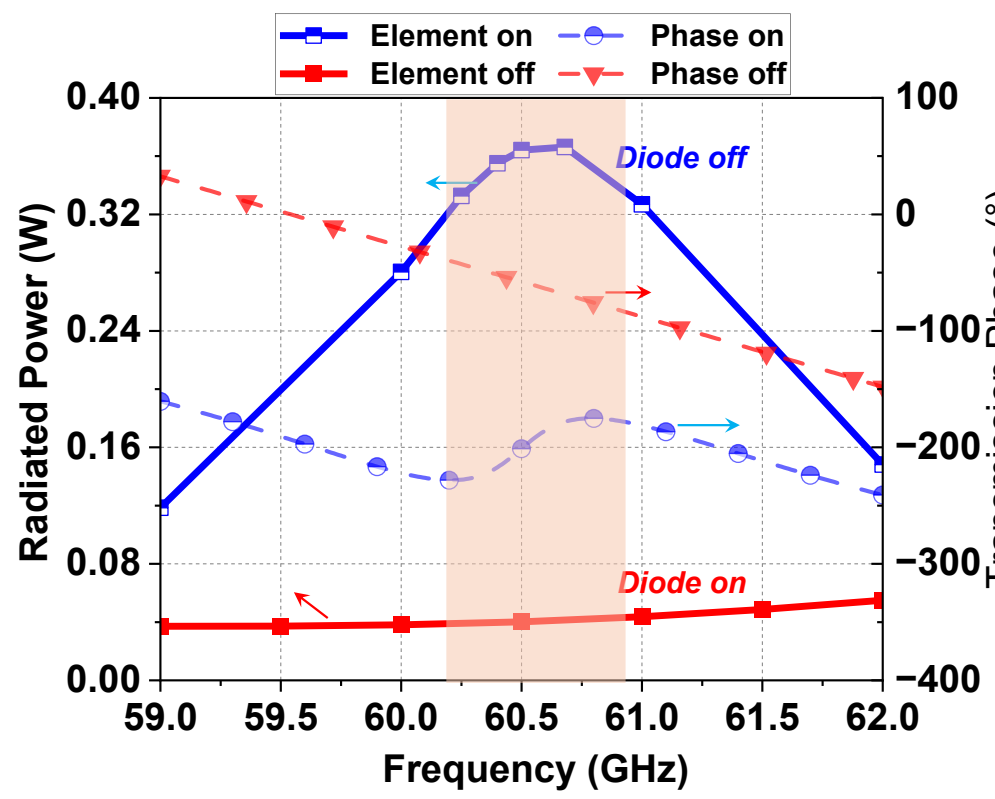

(b)

Fig. 2. (a) The meta-atom of the programmable DMA depicting radiation mechanism. (b) Radiated power profile during on and off states across the operating frequency band.

### B. *Current landscape of DMAs*

In recent years, DMAs have gained significant attention for their potential in next-generation wireless applications. DMAs are also referred to by other nomenclatures in the literature due to diverse wave manipulation phenomena, such as reconfigurable holographic surfaces [39], [40], and programmable/universal metasurface antenna [5]. More recently, some other architectures such as fluid antenna systems have been introduced to perform radiation pattern diversity and MIMO applications [41], [42]. While most existing studies on DMA-based communications are primarily limited to theoretical analyses [38], [43], their system model assumption is often made through an arbitrary number of meta-elements in simulation environments based on idealized microstrip apertures. Usually, these studies tend to overlook the impact of realistic planar waveguide-based hardware structures on radiation characteristics.

It is noticeable that the majority of the reported DMA designs to date (using varactor or PIN diodes) operate below 30 [4], [5], [31], [32], [41], [44]–[48]. Likewise, a DMA architecture based on liquid crystals has been demonstrated at 110 GHz [49]. However, despite these advancements on the antenna prototype level, a full-fledged end-to-end DMA-based wireless communication testbed has not been demonstrated to date at higher mmWave bands, such as 60 GHz and sub-THz. Although *reflective* coded metasurfaces have shown some testbed implementations below 30 GHz [7], [14], [18], nevertheless, the *radiative* DMA-based testbed implementation is still in its infancy. Therefore, beyond theoretical considerations and prototype-level designs, the development of system-level real-world DMA-assisted wireless communication platforms at higher frequencies remains a practical research challenge. This underscores the need for their system-level implementation and validation to realize their full potential in next-generation wireless networks.

### C. *Contribution and potential use-cases*

To the best of authors' knowledge, this is the first time a complete end-to-end 60 GHz DMA-based practical mmWave communication testbed is experimentally demonstrated. We experimentally demonstrate reliable and real-time over-the-air (OTA) transmission of a quadrature phase-shift keying (QPSK) modulated high-definition (HD) video signal through the DMA-assisted wireless link at a carrier frequency of 62 GHz. Utilizing a field programmable gate array (FPGA)-controlled high-speed digital signals, the proposed DMA-assisted setup can dynamically steer beams at a rate significantly higher than the symbol frequency. This enables efficient space diversity for multi-receiver communication, reducing outage probability and enhancing system reliability.

We present a compelling practical validation of a proof-of-concept DMA-assisted wireless communication testbed operating in the license-free 60 GHz industrial, scientific, and medical (ISM) band. Our previous work [50] proposed the fundamental prototype construction of the 60 GHz mmWave 1-D DMA and showed its dynamic beam patterns. While it provided the hardware foundation for potential applications involving dynamic beam-steering operations, it did not establish a complete end-to-end wireless communication system or experimentally validate real-time DMA-assisted communication capabilities in a 60 GHz mmWave system, as previously mentioned.

This work advances DMA technology from the prototype level to that of the real-world system-level application, and introduces new degrees of freedom, thereby enhancing the diversity and spatial multiplexing performance for future wireless communications. Notably, the conventional digital, analog, and hybrid beamforming architectures [51], which rely on time delay lines or power-hungry phase shifters are fundamentally redefined through DMA by eliminating these phase-shifting techniques. Instead, versatile beamforming is entirely performed by controlling the state of constituent meta-elements, enabling highly flexible beamforming operation. This transformation is particularly also advantageous in applications where size, weight, and power (SWaP) constraints are critical, such as in satellite communications.

In a wider application scenario, our contribution offers advanced reconfigurable beamforming antenna solution and its seamless connectivity with third-party 60 GHz mmWave connectorized modules. Notably, the license-free 60 GHz mmWave band enables localized high-capacity links and is expected to play a more prominent role in beyond 5G and 6G networks, such as in 60 GHz mmWave Wi-Fi venue hotspots, small cell wireless backhaul, fixed wireless access systems, license-free mmWave indoor healthcare monitoring and sensing systems, and integrated radar sensing [52]. Strategically placed DMA-based hotspots can significantly offload traffic from lower-frequency bands and macrocells, enhancing overall network performance [53].

Another potential solution offered by our DMA-assisted platform exists in next-generation ultra-reliable and low-latency mmWave industrial wireless communication[52]. These systems are envisioned as crucial enablers for meeting the stringent reliability and performance demands of Industry 4.0/5.0 use cases. [54]–[58]. Our proposed DMA-assisted testbed sets the foundation to support various indoor test scenarios for customized mmWave industrial wireless protocols, such as WirelessHP [59]. We envision that our 60 GHz DMA-assisted mmWave experimental testbed will provide a practical and scalable pathway to meet the stringent demands of beyond 5G and 6G communication systems while also serving as a catalyst for validating future sub-THz (such as D-band) 6G systems.

The rest of this paper is organized as follows. In Section II, we introduce the hardware design and radiation mechanism of DMA. In Section III, we elucidate the spatial dynamic beam-steering performance of the DMA through full-wave simulations and measurement results, and the beamforming control through FPGA. In Section IV, we thoroughly describe the implementation of DMA-assisted mmWave wireless communication setup, baseband modulation process, up/down-conversion to/from 60 GHz mmWave band, and experimental evaluation of the received video signal. Finally, the conclusion is presented in Section V.

## II. Hardware Design and Radiation Mechanism of DMA

### A. *Construction and Properties of Fully Addressable and Tunable Digital Meta-element*

We first realized a DMA element using a specially designed digital coding meta-element integrated with a substrate integrated waveguide (SIW) structure, shown in Fig. 2(a). A low-loss V-band SIW structure was designed to host the meta-element [50]. The meta-element is a complementary electric inductive-capacitive (CELC) resonant element that responds to the in-plane magnetic field component (dominant $TE_{10}$ mode) of the EM wave propagating in the SIW. A CELC resonator is a pure magnetic dipole resonant structure that couples and responds to the in-plane magnetic field, resulting in an absorbing profile at its resonant frequency.

The dipole moment representing a metamaterial element ($\boldsymbol{\mu}$) can be calculated as the product of the incident magnetic field ($\boldsymbol{H}$) and the element's polarizability ($\alpha$) [60]:

$$\boldsymbol{\mu} = \boldsymbol{H}\alpha \tag{1}$$

The magnetic polarizability of a CELC element manifests a Lorentzian shape curve as a function of the operating frequency ($\omega$), resonant frequency ($\omega_0$), coupling factor ($F$), and damping factor ($\gamma$) in the SIW structure:

$$\alpha = \frac{F\omega^2}{\omega_0^2 - \omega^2 + j\omega\gamma} \tag{2}$$

The full wave simulations of the CELC element in Fig. 2(b) reveal such Lorentzian resonance profile (bell-shaped with transmission phase reversal) of the CELC element during the coupled mode. The reference wave exciting the CELC has the dependence:

$$\boldsymbol{H_i} = H_0 e^{-j\beta x_i}\widehat{\boldsymbol{y}} \tag{3}$$

where $H_0$ is the initial magnetic field flowing into the SIW structure, $\beta$ is the effective waveguide propagation constant in SIW (given as $\beta = \frac{2\pi}{\lambda}$).

To achieve fully addressable 1-bit ON-OFF spatial modulation of the CELC element, it is integrated with two PIN diodes (two diodes maintain the symmetry and suppress the radiation well in the element-off state). Since a PIN diode only has two states (ON/OFF), the meta-element loaded with PIN diodes achieves binary discrete amplitude control. The switching states of diodes are controlled through a carefully designed DC biasing network interfaced with the FPGA, ensuring precise modulation between the ON and OFF states [50].

An arrangement of such tunable meta-atoms in the SIW structure with intelligently designed biasing network leads to DMA array formation. By utilizing different biasing states (coding combinations) of PIN diodes controlled through FPGA programming, the phase difference between the radiating elements can be dynamically changed, thereby realizing versatile antenna-pattern generation and reconfiguration in real-time.

The far-field beam pattern of the constituent CELC meta-element at a distance ($r$) from the source is given by the far-field magnetic field vector $\boldsymbol{H}$ as [30]:

$$\boldsymbol{H} = \frac{\omega^2 m}{4\pi r} cos\theta[e^{-jkr+j\omega t}]\widehat{\boldsymbol{\theta}} \tag{4}$$

where $r$ is the approximated magnitude of the difference between the location of the radiation source and the observation point at a far-field distance, $\widehat{\boldsymbol{\theta}}$ is the unit vector in the direction of the radiated wave, and $k$ is the free-space wave number ($k = 2\pi/\lambda_0$). In the array topology, these metamaterial elements can be treated as selecting the excitation wave (also termed as reference wave) at their locations on the SIW aperture by $x_n = nd$, where $n$ is an integer and $d$ is the spacing between any two adjacent CELC elements. Then the far-field radiation pattern from the 1D (one dimensional) DMA can be approximated by superimposing the fields sourced by all of the elements as:

$$\boldsymbol{H_{rad}} = H_0 \frac{\omega^2}{4\pi r} e^{-jkr} cos\theta \sum_1^N \alpha_n(\omega)\, e^{-jx_n(\beta+ksin\phi)}\, \hat{\boldsymbol{\theta}} \quad (5)$$

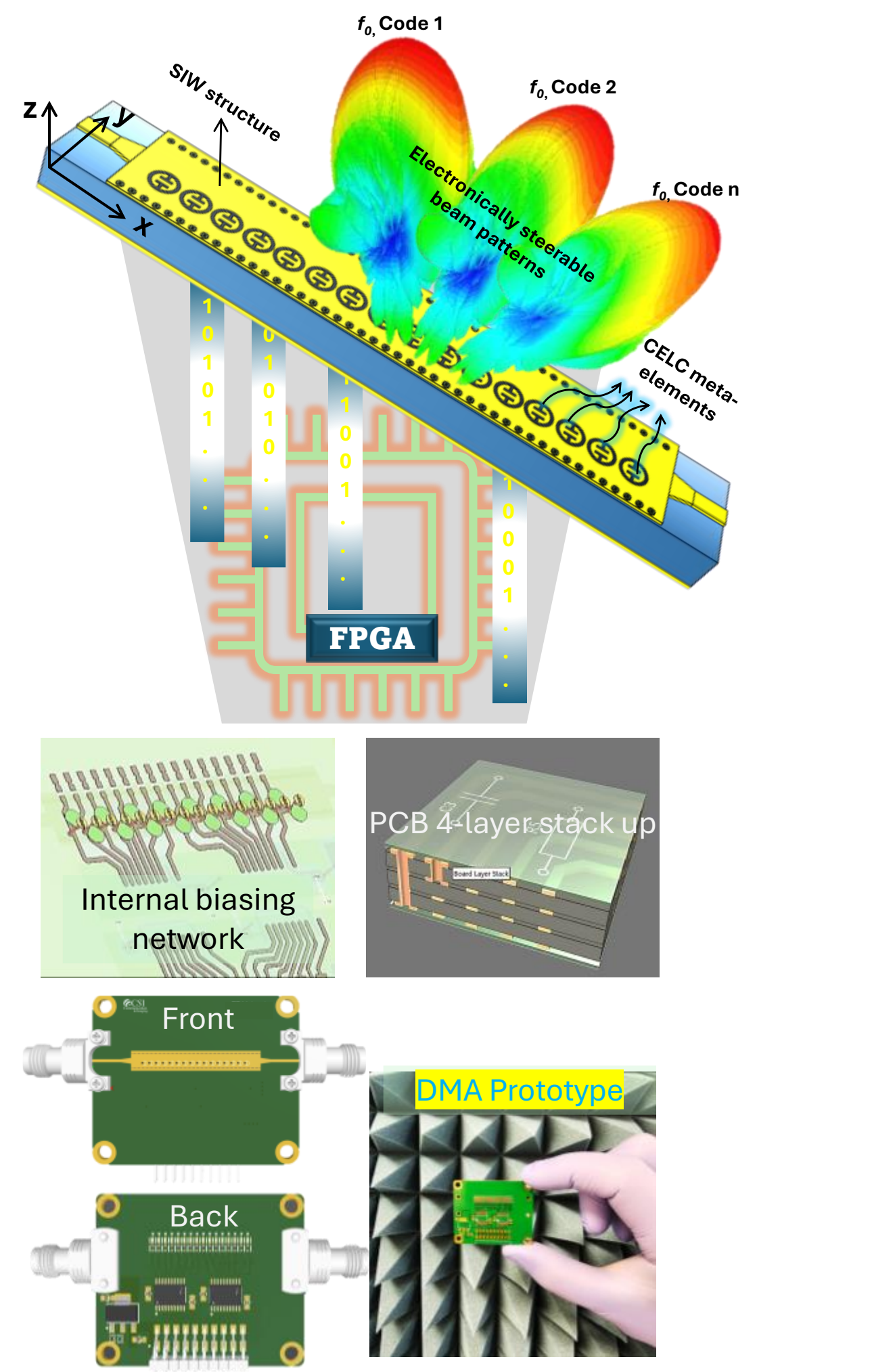

Fig. 3. Proposed schematic diagram of 1-D DMA along with fabricated prototype, internal biasing circuitry, and PCB layer stack-up view.

Here $cos\theta \sum_1^N \alpha_n(\omega)\, e^{-jx_n(\beta+ksin\phi)}$ is the array factor of 1-D DMA, with $\phi$ being the angle of propagation normal to the surface of the 1D DMA surface.

### *B. Radiation Mechanism of DMA*

Figure 2(a) illustrates the radiation mechanism of the CELC element, demonstrating how the coding meta-element interacts with EM waves under different digital states, resulting in distinct electromagnetic responses. The induced surface currents reveal physical digital coding states ‘0’ and ‘1’, (practically zero voltage will show the diode ‘OFF’, and meta-element ‘ON’ or ‘state 1’; whereas 1.33 V switches the diode ‘ON’, meta-element ‘OFF’ or ‘state 0’). To put this into physical context, when the diode is switched on, the CELC element is effectively shorted with the outer SIW structure. In this state, it loses radiation properties and does not couple with the waveguide mode (as it becomes analogous to the part of a transmission line). Conversely, when the diode is switched off, the CELC element is effectively isolated from the outer SIW structure, presenting an open circuit. It then couples the portion of the wave through the gap complement of the SIW, and radiates it into the free space at its designed operating frequency.

It is worth mentioning here that such radiation phenomenon of DMA makes them energy efficient because the CELC elements manifest their antenna (radiative) properties when the diode is switched off. Consequently, PIN diodes consume negligible current in their off state, significantly reducing the overall power consumption of the DMA.

The PIN diode (MADP-000907-14020W) has fast switching speed of 2–3 ns and was modelled as a lumped circuit element with series resistor-inductor-capacitor (RLC) circuit parameters R = 5.2 Ω, L = 30 pH, and C = 0 pF for the ‘diode ON’ state and R = 10 kΩ, L = 30 pH, and C = 25 fF for the ‘diode OFF’ state. The radiated power profile of the CELC meta-atom during on and off states is shown in Fig. 2(b). Notably, the designed CELC element radiates around 9 times more than that of its non-radiation state at 60.6 GHz, which confirms a 1-bit response.

### *C. Implemented 1-D DMA Hardware Prototype*

As shown in Fig. 3, the implemented 4-layer DMA prototype consists of a 1-D array of 16 digitally controllable elements and has a total size of 4.36 cm × 3.6 cm × 0.154 cm. Further detailed dimensions are provided in [50]. The RF metasurface layer was designed using a high-performance Rogers 3003 substrate, with a dielectric constant of 3, thickness of 0.25 mm, and copper cladding of 17.5 microns. The DMA is fed through a 1.85 mm standard connector from one side of the SIW. The reference wave generated from the feed propagates to each DMA element and can radiate energy to the free space. The residual energy of the EM waves is absorbed by a high frequency 50 Ω matched load impedance attached to the other end of the SIW.

The radiation amplitude of each DMA element is controlled through FPGA (Digilent Arty S7-25T) in which a custom-made software program was pre-loaded to control the binary coded sequence. The DMA design consists of 4 layers where the top layer includes meta-elements, second layer is RF ground, while third layer is dedicated to the DC biasing and isolating network, as shown in Fig. 3. The DC output of FPGA is +3.3 V which is converted to +1.33 V using a voltage regulator IC embedded on the bottom layer for switching PIN diodes. The biasing network for all PIN diodes was designed in a parallelized manner. This means that the digitally coded signal from FPGA is simultaneously provided to all meta-elements which significantly reduces the beamforming switching time. In the case where all PIN diodes are OFF, the meta-elements would be in a radiating state with minimal power consumption and pointing the main beam direction towards the broadside (nearly 0°) at 62 GHz.

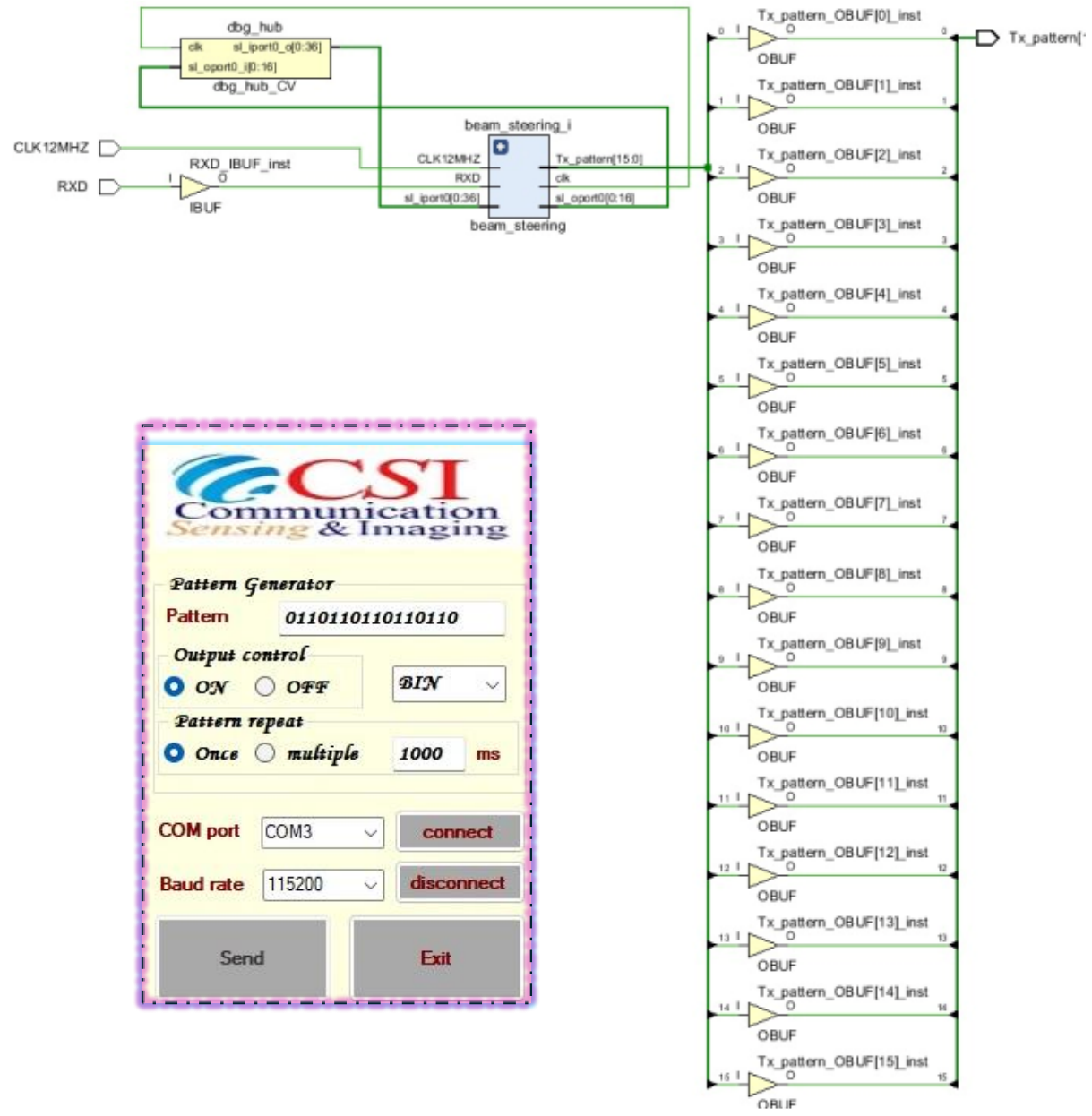

Fig. 4. Schematic diagram of the parallelized FPGA control scheme, and the customized GUI.

However, this scenario represents a trivial case and does not fully exploit the reconfigurability of the DMA. The operating region of the DMA is mainly between 60 GHz to 62 GHz.

## III. IMPLEMENTED DYNAMIC BEAM-STEERING

The beam-steering mechanism of the proposed DMA is fundamentally based on the phase accumulation of the guided wave, which sequentially excites the meta-elements along the aperture with the necessary phase variation to form directional beams. By employing different diode tuning schemes from FPGA, the effective phase difference between radiating elements is dynamically changed, enabling the generation of diverse radiation patterns.

### *A. Digital Control through FPGA Programming*

We controlled dynamic beam-steering by simultaneously controlling the radiation state of 16 CELC elements using an FPGA. The FPGA includes a synchronization subcircuit that generates clocks and sync signals while receiving azimuth control inputs from the graphical user interface (GUI). It can reconfigure beam steering within 10 ns using a 100 MHz clock frequency.

The FPGA receives steering codes in binary format via the UART (universal asynchronous receiver-transmitter) protocol, along with control messages that configure transmission parameters, including switching beam steering on/off, selection of single or multiple codes, and defining the switching time between different codes. The schematic design of FPGA control scheme is presented in Fig. 4. The program is developed in VHDL using Xilinx Vivado software.

We designed a user-friendly custom-made GUI to facilitate steering code transmission in multiple formats, including binary, hexadecimal, and decimal. The GUI was designed using MS Visual Basic and a snapshot is shown in Fig. 4. The GUI allows users to input control codes, toggle beam steering on/off, switch between single and multiple azimuth steering modes, as well as adjust the timing between code switches according to symbol update rates.

Table I. METASURFACE BEAMFORMING CODES FOR DIRECTIONAL AND MULTIPLE BEAMS.

| Code-book | Coding sequence | No. of Beams | MLD°/ (Gain (dBi)) | HPBW° |
|---|---|---|---|---|
| Code 1 | 1001001001001001 | 2 | -49°, +23° 11.12, 11.85 | 13°, 11° |
| Code 2 | 0001000100010001 | 2 | -16°, +31° (7.97, 7.95) | 10°, 17.5° |
| Code 3 | 1010101010101010 | 1 | -9° (8.53) | 8° |
| Code 4 | 1001100110011001 | 1 | -8° (8.5) | 24° |
| Code 5 | 1010101010000000 | 1 | +36° (10.2) | 13.5° |

### *B. Dynamic Beam-steering Performance of DMA*

For a 16-element linear DMA, the possible coding space contains $2^{16}$=65,536 binary-coded combinations. Among these, many configurations produce narrow beams, while others can generate wider beams or even multi-beam patterns. This versatility is the hallmark of DMA, enabling it for flexible beamforming capabilities. However, it should be noted that from a wireless communication perspective, many of these codes may not meet performance requirements if they result in high side lobe levels or exhibit unmatched input impedance.

For proof of concept, here we utilize two sets of coding sequences at 62 GHz, one for generating a single directional beam, while the other for producing dual beam. The corresponding binary coding combinations are presented in Table I in which the binary state corresponds to the radiation state of the meta-elements. The inverse of these coding sequences was applied through the FPGA, as the forward-biased state of the PIN diode corresponds to state-1 (high voltage) in FPGA program. Similarly, in simulations, the diode switching states were implemented on the meta-elements of the DMA's top RF layer using full-wave analysis in CST Microwave Studio, following the above specified operating bias conditions of the diode model.

As shown in Fig. 5, code 1 and code 2 produce dual beams that can be utilized for point-to-multipoint (P2MP) multi-user links. Code 3, code 4, and code 5 generate directional single beams for point-to-point (P2P) single user links towards -9°, -8°, and +36° respectively. Furthermore, the half-power beamwidth (HPBW), main lobe direction (MLD) and the corresponding gain values of such codes are given in

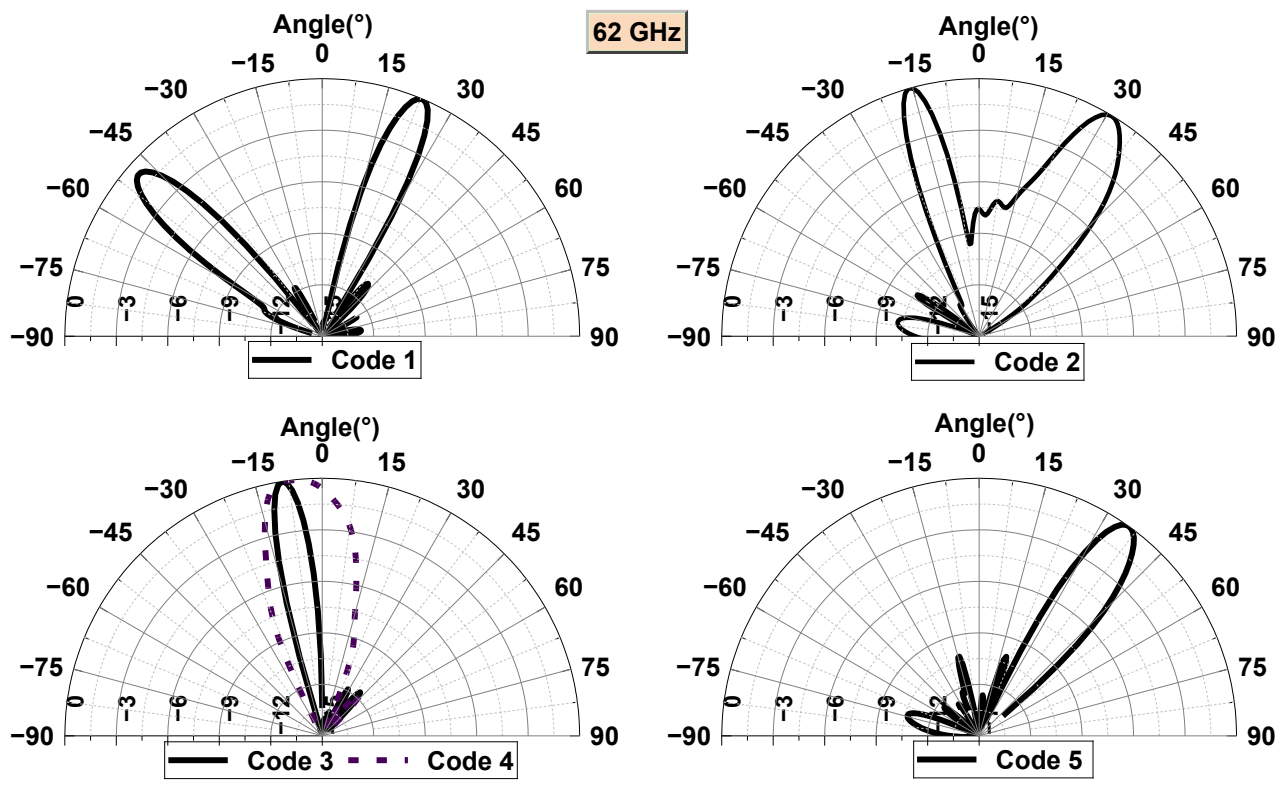

Fig. 5. Simulated beam patterns with different beamforming coding combinations at 62 GHz.

Table I. The simulated beams are validated through measurements using the far-field method as explained in [50]. The MLD of the measured radiation patterns is in good agreement with the simulated results, as presented in Fig. 6(a). Note that for the proof-of-concept demonstration in this work, two such beams are sufficient to validate our testbed. Nevertheless, the DMA can generate various other beams (narrow, wider, dual and tri beams) using versatile binary-coded sequences programmed through FPGA. The -10 dB input impedance (reflection coefficient) for code 1 and code 3 remains below -10 dB, indicating efficient input power acceptance of DMA for these coding sequences around 62 GHz, as shown in Fig. 6(b).

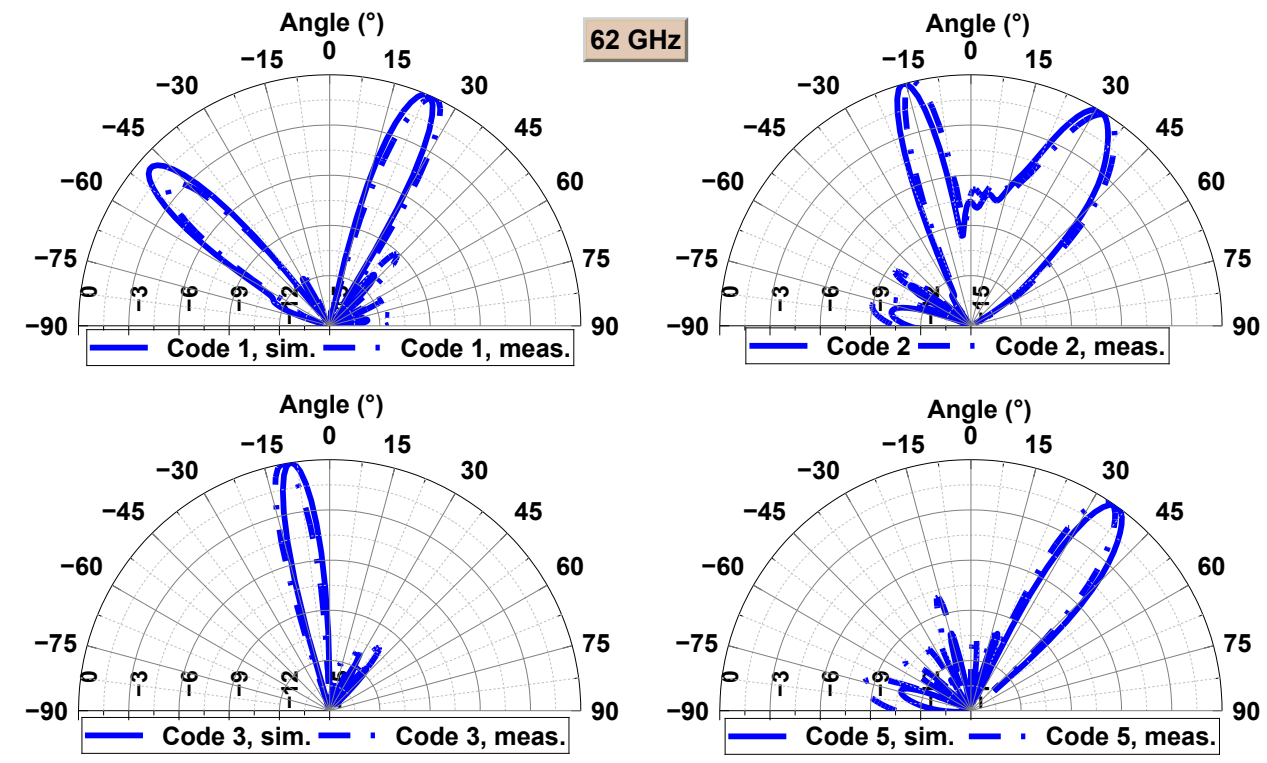

(a)

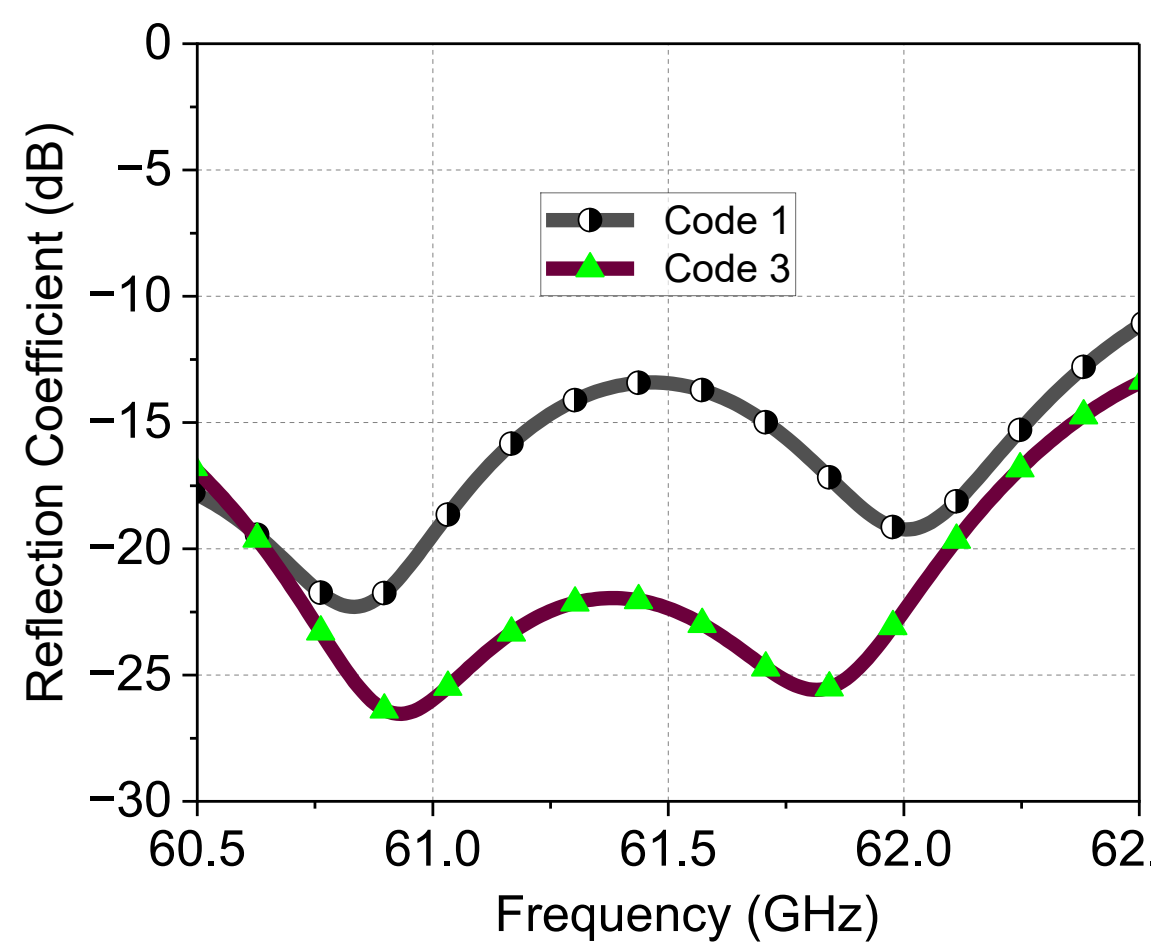

(b)

Fig. 6. (a) Measured beam patterns for four different beamforming coding combinations at 62 GHz. (b) Input impedance of code 1 and code 3, below -10 dB depicts good antenna matching in the waveguide structure.

## IV. Implementation and Evaluation of DMA-Assisted Communication Platform

### A. *Implementation of 60 GHz mmWave Test Setup*

The implemented testbed consists of various stages and hardware modules on the transmitter (Tx) and receiver (Rx) sides, as demonstrated in Fig. 7. The description of these modules is explained below.

***Transmitter-Side Setup:*** The Tx side comprises a mmWave programmable DMA prototype, an FPGA to generate binary sequences with dedicated GUI, a universal software radio peripheral (USRP, a commercial software-defined radio (SDR) transceiver) for baseband signal processing and RF signal source. A host personal computer (PC) was used to control the parameters of Tx USRP B200 as the source of data bits through the GNU Radio software development kit. The Tx USRP performs baseband signal processing and RF modulation (to generate a QPSK-modulated video stream) with carrier frequency centered at 1 GHz. The Tx host computer can control the parameters such as the carrier frequency, gain, sampling rate, and modulation mode. The 1 GHz modulated carrier signal is transmitted to the mmWave module for up-conversion using a phase-matched SMA-to-MCX cable at the baseband interface of the mmWave module.

For mmWave up-conversion, we used Analog Devices (ADI) HMC6350 evaluation board which consists of an HMC6300 transmitter integrated circuit (IC) that provides a mmWave signal within the range 57–64 GHz. The Tx IC has an integrated frequency synthesizer that creates a low phase noise local oscillator (LO) signal between 16.3 GHz and 18.3 GHz, with a step size of 250 MHz. The synthesized LO is divided by two to produce an intermediate frequency (IF) range of 8 GHz to 9.1 GHz. The IF signal is filtered and amplified with a variable gain of 14 dB before it is mixed with three times the LO frequency, upconverting it to an RF frequency range of 57 GHz to 64 GHz. We controlled the mmWave module's Tx parameters such as RF attenuation, IF attenuation, and mmWave frequency, using its dedicated GUI, accessed through the module's USB interface.

The output of mmWave module is connected to the DMA input using an MMPX-to-1.85 mm standard cable. The DMA then transmits this mmWave signal over the air.

***Receiver-Side Setup:*** The Rx setup mainly consists of a down-conversion module and a Rx USRP for demodulation.

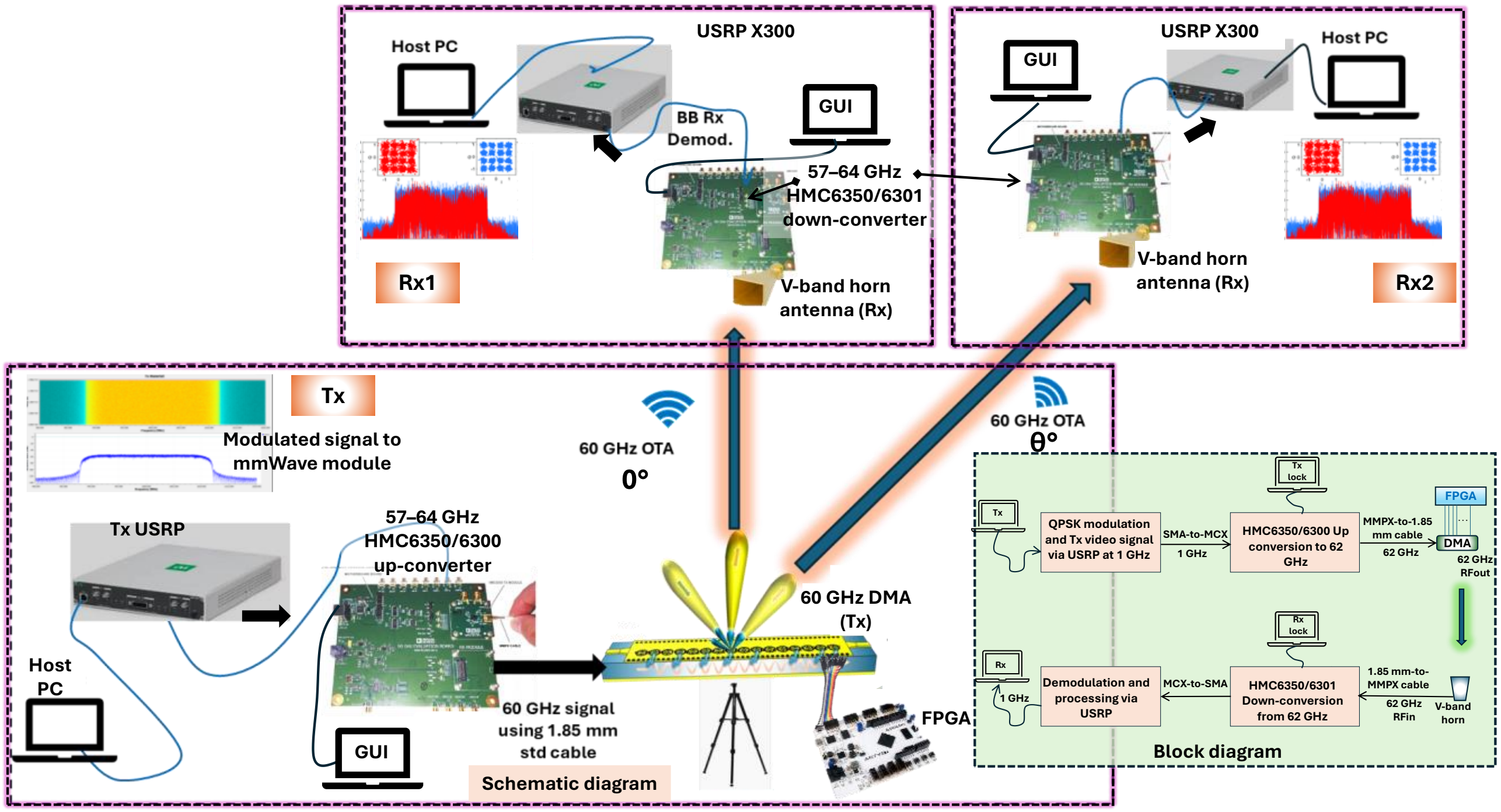

Fig. 7. Conceptual schematic diagram of the mmWave DMA-assisted wireless communication system. By applying different coded sequences from the FPGA to PIN diodes, the overall phase distribution on the DMA aperture can be manipulated which is used to encode the signal information directly and transmit different coding streams to multiple designated users at different locations simultaneously and independently.

A V-band linear polarized horn antenna with 15.5 dBi gain was used to capture the mmWave signal. It was connected to HMC6301 Rx module integrated on a separate HMC6350 board using a 1.85 mm-to-MMPX cable.

The HMC6301Rx module has an integrated frequency synthesizer that generates a low phase noise LO between 16.3 GHz and 18.3 GHz. As any of the 57 GHz to 64 GHz signal enters the module through its integrated low-noise amplifier input, the LO is multiplied by three and mixed with the LNA output with a gain of 20 dB. It is then down-converted to an 8.14 GHz to 9.1 GHz sliding IF signal with variable gain of 12 dB. This IF signal is then down-converted by mixing with the synthesized LO divided by two.

The down-converted signal from the mmWave module is then provided to the input receiving port of USRP X300 using an MCX-to-SMA cable. The baseband signal processing and demodulation is performed at Rx USRP which was connected to the host PC through a crossover Ethernet cable. SDRangel software was used to analyze the received signal and real-time video reception.

### *B. Experimental Evaluation of Real-time Communication Testbed*

We implemented the digital video broadcast (DVB) standard for video data transmission. The transmitted signal first passes into multiplexer adaptation and energy dispersal, then encoded by a serial concatenated coder which is a concatenation of an inner coder and an outer coder to control errors. Specifically, the Reed-Solomon coder and the convolutional coder are performed as the outer coder and inner coder, respectively, and connected via an outer interleaver to separate bursts of errors. After that, the encoded signal is processed by a bit and symbol inner interleaving and bit interleaved baseband frames are then treated by the DVB constellation mapping to form the QPSK modulated signal with a code rate of 5/6. Finally, the modulated signal is passed through a decimating filter using the fast convolution method through Fast Fourier Transform (FFT). The block diagram of DVB signal scheme is depicted in Fig. 8. The modulated carrier frequency was set at 1 GHz, and the symbol rate was set to 2 MSymbols/s at Tx USRP. This signal was fed to the input of HMC6300 up-conversion module. The DMA-assisted communication platform was deployed in an open lab environment to evaluate real-world wireless communication scenario, as shown in Fig. 9.

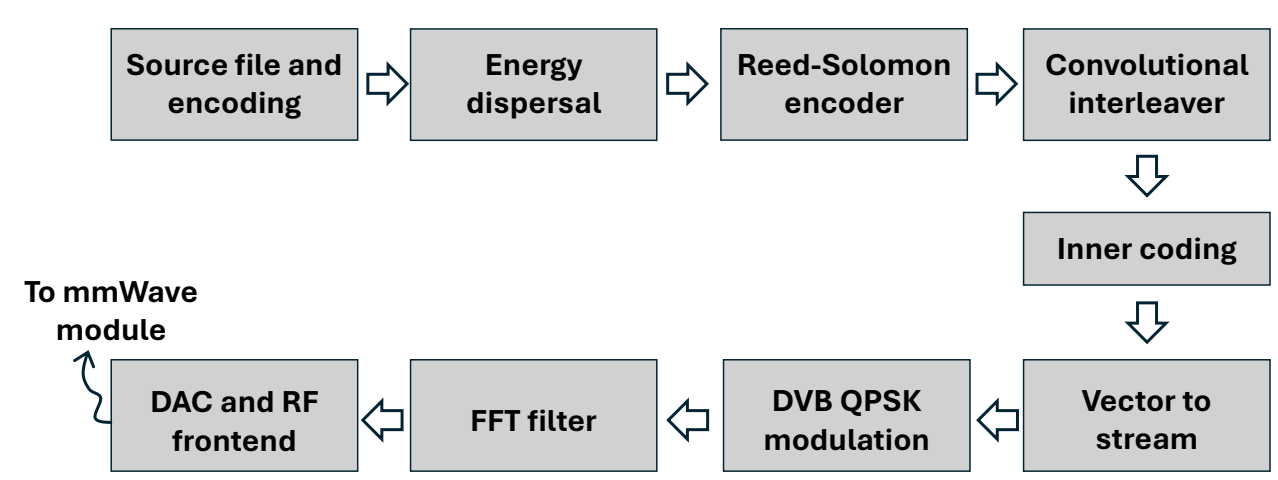

Fig. 8. Functional block diagram of DVB QPSK modulated Tx signal scheme.

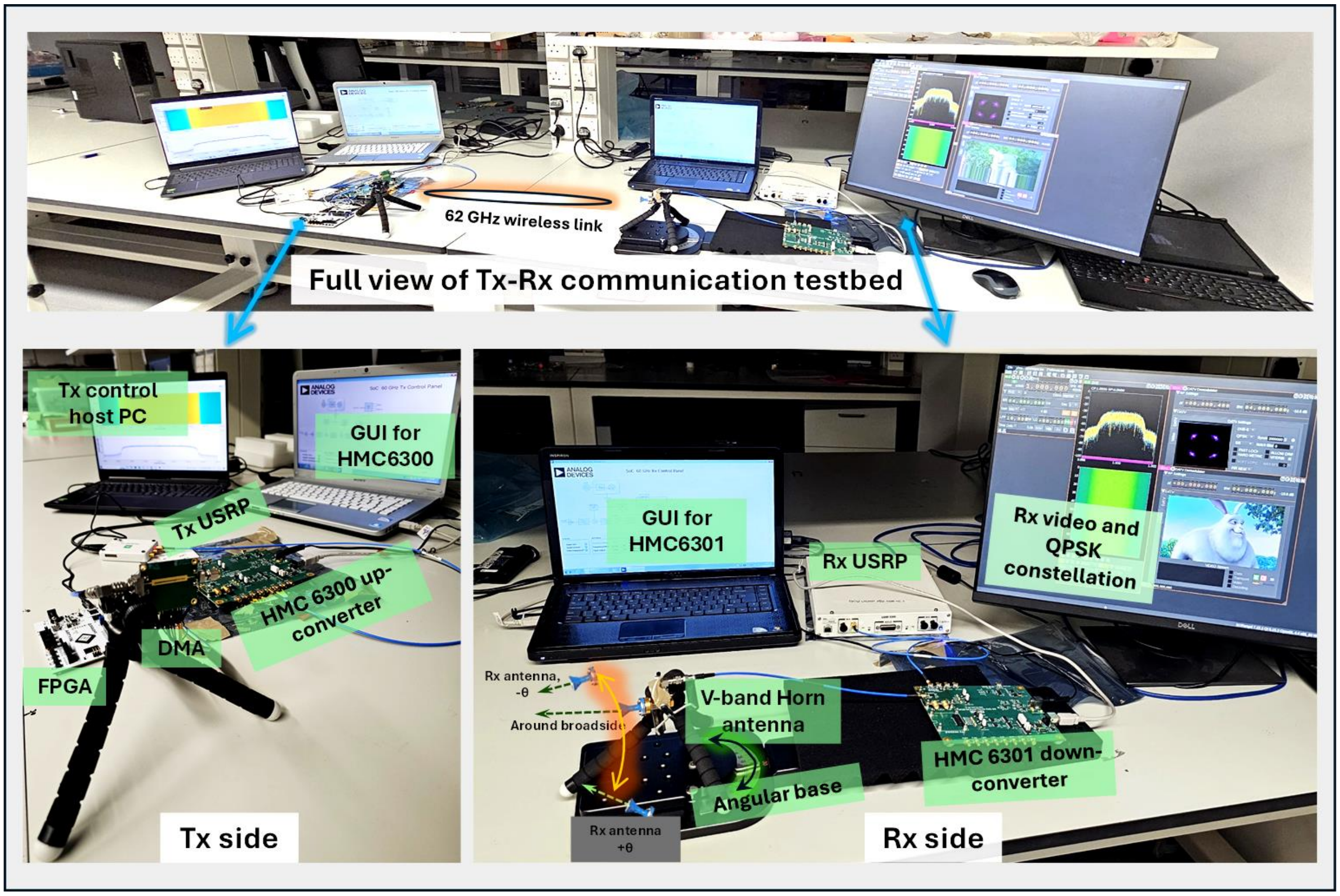

Fig. 9. Experimental testbed setup for mmWave wireless communication based on DMA. A QPSK-modulated video signal is transmitted through DMA (shown bottom left Tx terminal) to the receiving end (shown towards the bottom right) in real-time with 1280 × 720 resolution, over mmWave carrier frequency of 62 GHz.

We configured the DMA for two use cases. First, by applying code 3 from FPGA, we generated a single beam from DMA to serve a single channel wireless link. In the second experiment, we used code 1 to generate dual beams to serve two users simultaneously at an angular separation. The distance between the Tx DMA and Rx horn antenna was set to 1 meter. As the DMA was positioned to produce narrow beam coverage in the azimuth (fan-shaped), the antennas were aligned using a laser beam.

First, the modulated Tx signal was examined using a real-time spectrum analyzer (Siglent SSA 3075X-R) to verify the waveform shape and ensure a safe RF power level before being fed into the mmWave module, as demonstrated in Fig. 10 (top). The 1 GHz modulated signal was fed to the HMC6300 mmWave module and its frequency was locked at 62 GHz with a transmit power of +7 dBm.

On the Rx side, OTA mmWave signal was received by a V-band horn antenna. The HMC6301 receiver IC was locked at 62 GHz using its GUI. The Rx horn antenna was aligned for best constellation recovery through laser beam alignment. The down-converted signal was fed to Rx USRP. The gain of the modulated signal was fine-tuned using the GUI of the Tx USRP to prevent receiver saturation and enhance the quality of received constellation, based on real-time observations of the Rx waveform. The real-time HD video stream was received and analyzed through SDRangel software on Rx host PC. The important parameters such as the received signal waveform, QPSK constellation points, and real-time video snapshot are displayed in Fig. 10 (bottom). The real-time video transmission was received with a resolution of 1280 × 720 and an average transmission rate of 3.2 Mbps. As shown in the constellation diagram, the four points are equally spaced around a circle, implying that the received signal is demodulated successfully. This validates that our DMA-assisted communication platform supports real-time transmission of HD video and data.

In our experimental setup, we tested the link at a nominal distance ($d_0$) of 1 meter and extended it to 1.6 meters, where reliable constellation recovery was achieved. The achievable range ($d$) in our setup is primarily constrained by the path loss, transmit power $P_T$, receiver sensitivity, and the aperture gain of the Tx DMA ($G_T$) and Rx horn antenna ($G_R$). The free space path loss at a 1 m reference distance can be calculated as (5):

$$PL_{d0} = 20\log_{10}(4\pi d_0/\lambda) \quad (5)$$

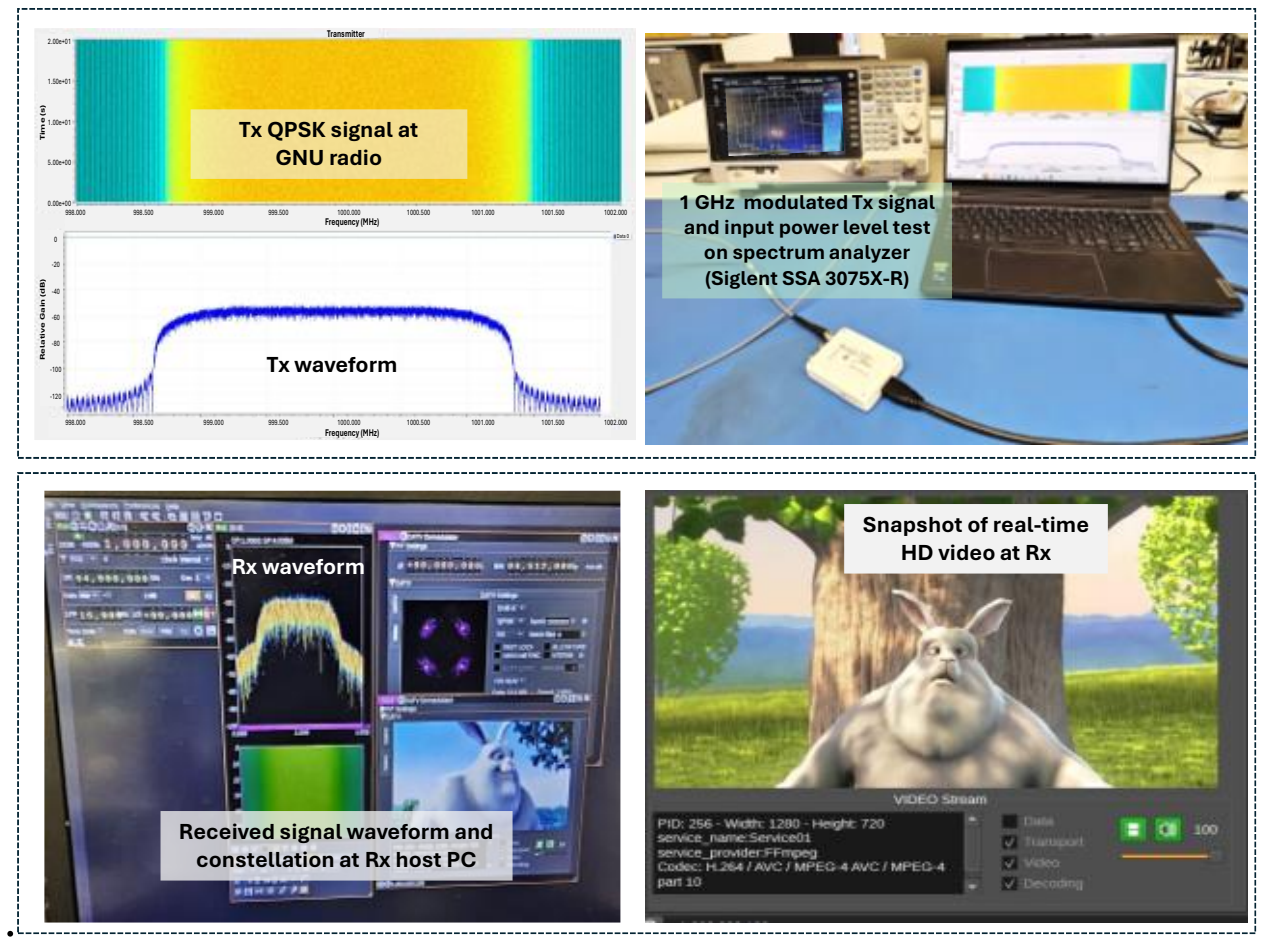

Fig. 10. (Top) experimental validation of Tx signal on spectrum analyzer. (bottom) graphical interface of received signa profile at the Rx USRP revealing spectrum, constellation, and real-time video snapshot.

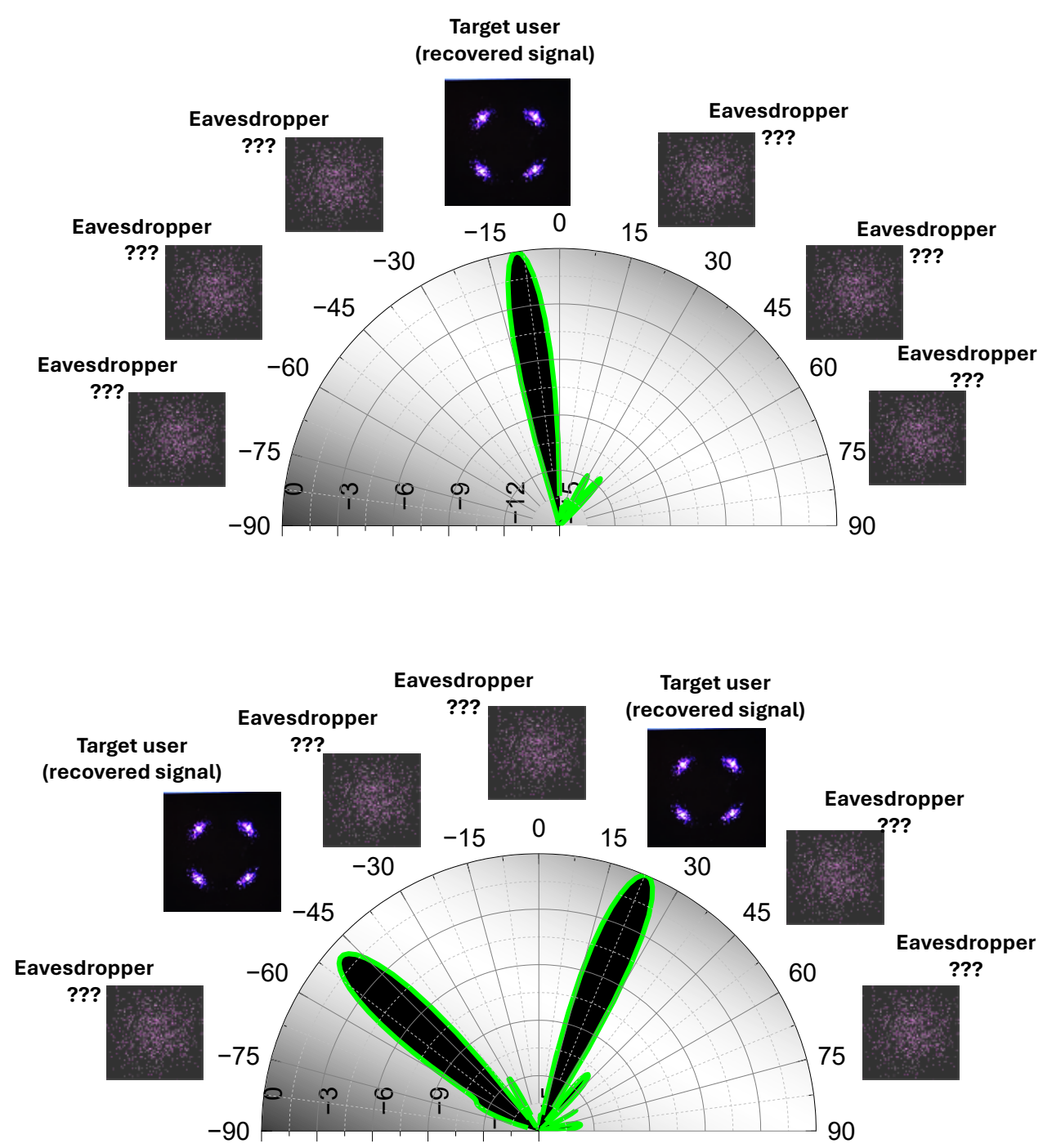

Fig. 11. Experimental decoded constellation diagrams at different receiver directions (code 3 for top, and code 1 for bottom figure) at 62 GHz. This validates the inherent security of DMA by directional beams at the fundamental frequency. Only the intended user(s) within the main beam can successfully decode the transmitted information with the correct constellation, while eavesdroppers at other locations are unable to recover it, enhancing PHY security.

where $\lambda$ is the wavelength. The average path loss over a separation distance $d$ between Tx and Rx can be calculated by the path loss exponent model, given in (6) as:

$$PL_d = PL_{d0} + 10n \log_{10}(d/d_0) \tag{6}$$

The path loss exponent ($n$) characterizes how fast the path loss increases with the increase of Tx-Rx separation. Its value represents the signal power loss from the transmitter antenna to the receiver antenna, and does not depend on the antenna gains or the transmitted power levels. Nevertheless, in an indoor environment, the value of $n$ can be taken as 2. The calculated path loss in the experiment setup is 68.28 dB at a 1-meter distance and a carrier frequency of 62 GHz.

Table II. KEY PARAMETERS OF THE WIRELESS COMMUNICATION TESTBED.

| Parameter | Value |
|---|---|
| Baseband (Low) frequency at Tx and Rx USRPs | 1 GHz |
| RF Power From Tx USRP to HMC6350 module | -25 dBm |
| OTA mmWave frequency | 62 GHz |
| OTA mmWave output Tx power | +7 dBm |
| Tx DMA gain | 8 to 10 dBi |
| Tx-Rx distance | 1 m |
| Path loss | 68.28 dB |
| Rx horn antenna gain | 15.5 dBi |
| Rx gain at LNA and IF stage of HMC6301 | 32 dB |
| Modulation scheme | QPSK |
| Code rate | 5/6 |
| Symbol rate | 2 MS/s |
| Achieved average transmission rate | 3.2 Mbps |

The received power ($P_R$) during the communication link is governed by Friis transmission equation [61] and is based on the transmit power ($P_T$), antenna gains, and path loss as (7):

$$P_R = P_T + G_T + G_R - PL_d \tag{7}$$

It is evident from (7), that future DMA designs with 2D planar geometry and higher gain can enhance the link range, while keeping the Tx power constraint in limits. Other key parameters of the communication testbed are listed in Table II.

In order to test multi-user links through DMA, we generated code 1 from FPGA to produce two beams towards -49° and +23° (as per simulations and measured patterns) to serve two independent receivers simultaneously.

Because of unavailability of the another down-conversion mmWave module, we analyzed and retrieved the real-time video for two channels one by one through aligning the Rx horn antenna at directions $\theta = -49°$ and +23°, with respect to the normal of the DMA. It was experimentally examined that the received video streams demodulated at these two different locations are same as the transmitted one. Moreover, when the receiver antenna location was changed (say Rx positioned at $\theta$ = -15° or +12°), the synthesized signals from the two physical channels cannot be correctly demodulated, even with a significant increase in Tx gain, as demonstrated in Fig. 11.

A key advantage of the DMA-assisted mmWave system is its ability to enhance physical layer (PHY) security. Through precisely controlling the directional radiation pattern, the transmitted signal can be steered toward the intended user while significantly degrading reception for unintended users (or eavesdroppers). As demonstrated in Fig. 11, the measured radiation pattern and decoded constellation diagrams confirm that only the receiver within the main beam direction can successfully decode the transmitted information, while eavesdroppers positioned outside the beam experience severe signal degradation, rendering the constellation unintelligible. Such directional spatial beamforming provides a fundamental mmWave PHY security.

### C. Potential Avenues of DMA-assisted Wireless Systems

***PHY Security:*** Besides directional spatial beamforming, as implemented in this work, other sophisticated techniques for PHY security can be employed through DMA-based system. These techniques include direct information modulation and chaotic information metasurface approach, which involve high entropy, non-repeating chaotic sequences [15], [20]. These support direct PHY secure communication whereby the information itself is encoded in the spatio-temporal characteristics of the electromagnetic wavefronts and a purposeful concealing methodology for non-targeted receivers. Such an approach does not rely on standard decryption operations, as the intended receivers to which the information can be reconstructed will always have the opportunity to recreate the original stream, but the unintended receivers obtain only random, incomprehensible transmissions. Thus, inferring from these concepts, a DMA-based setup can yield similar chaotic facilitation for PHY security since fine control can be applied to every single meta-atom of DMA through FPGA.

***Space-time coding and ISAC:*** ISAC is one of the demanding research topics under the ambit of 6G, for which DMA-enabled ISAC systems can reduce overall hardware complexity and power requirements. DMAs act as transceivers (instead of mere reflective surfaces) and can enable time-division multiplexed support for both sensing and communication. For instance, independent beams can be dynamically formed, one dedicated to communication and another to sensing. Although such DMA-based holographic spatial beamforming schemes can enable multi-beam implementations for simplistic ISAC applications [40], even more exciting EM wave phenomena can be realized by integrating time modulation with spatial coding, leading to the emergence of space-time coding capabilities. It is worth mentioning here that the space-time wave manipulation has been widely achieved in recent years through coding reflective metasurfaces [6], [7], [12], [25], [27], however, implementation with DMA-based systems at mmWave bands is relatively new to explore. Time-varying control signals can be generated from an FPGA with spatial variation, allowing directional beams at the carrier frequency as well as its harmonics. Such spatio-temporal control offers a greater degree of freedom to devise simultaneous communication and sensing channels. As such, by exploiting carrier frequency for communication link and harmonic components for sensing, they hold the potential to eliminate the need for separate sensing modules or dedicated control links from the base station, thereby allowing sensing and communication to occur concurrently and without mutual interference. Such frequency-division-based spatiotemporal encoding holds the immense potential for practical implementation of DMA-assisted mmWave ISAC, combining communication and sensing on a single shared hardware platform. Furthermore, spatio-temporal control can create a synergy of DMA-assisted RIS-enabled wireless paradigm, as elucidated in the introduction section and illustrated in Fig. 1b.

While the current 1D DMA-based testbed has been experimentally validated for spatial beamforming at the carrier frequency for communication, it serves as a scalable hardware platform that can be extended to planar (2D) architectures supporting advanced space-time modulation and 3D beamforming. In space-time modulation, the radiation state of individual DMA elements can be fully addressed dynamically varied over time, enabled by parallelized control signals from the integrated FPGA. This temporal modulation induces frequency harmonics around the carrier frequency $f_c$, resulting in harmonic components $f_n = f_c \pm nf_m$, where $f_m$ is the modulation frequency and $n$ is the harmonic order. For example, at a carrier frequency of 62 GHz with an FPGA-driven modulation frequency of 50 MHz, first harmonic signals can be generated at 61.95 GHz and 62.05 GHz, which can be further manipulated for multi-frequency beam steering or dual-functional sensing and communication. Such implementation would require enhancements in the update rate of the control logic from the FPGA, switching speed, duty cycle, and the temporal resolution of the biasing circuitry. The full implementation of this concept is beyond the scope of this work and is left for future exploration.

## V. Conclusion

In this paper, we present a proof-of-concept practical implementation of a DMA-based mmWave wireless communication testbed, demonstrating real-world operation at 62 GHz. The hardware design of the mmWave DMA with software controllable radiation patterns was elucidated, alongside the implementation of amplitude-controlled beamforming through a high-speed FPGA. The proposed testbed integrates a DMA-based setup with an HMC6350 motherboard and HMC6300/6301 up/down-conversion modules for high-frequency mmWave signal transmission and reception. Additionally, it utilizes commercial USRPs for low-frequency signal generation and processing, providing a robust framework for mmWave wireless communication. To validate the DMA-assisted testbed, a QPSK-modulated HD video signal was successfully transmitted over-the-air at 62 GHz via the DMA, both over a single-user link and simultaneously across two independent physical channels to separate legitimate receivers, ensuring accurate demodulation and retrieval at each receiver. The spatial diversity of DMA-generated beams was tested for PHY security by restricting signal decoding to intended users within the main beam, while significantly degrading signal quality for potential eavesdroppers. In the future, we aim to unlock the full potential of DMA-assisted systems for emerging next-generation use cases, including ISAC, and dynamic spatio-temporal modulation through space-time coding techniques at mmWave bands, enabling secure and encrypted multi-user indoor wireless links.

## Acknowledgment

This work was supported by CHEDDAR: Communications Hub for Empowering Distributed ClouD Computing Applications and Research, funded by the UK EPSRC under grant numbers EP/Y037421/1, EP/X040518/1, and by the UK EPSRC grant number EP/Z533609/1.

## References

[1] N. Shlezinger, G. C. Alexandropoulos, M. F. Imani, Y. C. Eldar, and D. R. Smith, "Dynamic Metasurface Antennas for 6G Extreme Massive MIMO Communications," *IEEE Wireless Communications*, vol. 28, no. 2, pp. 106–113, 2021, doi: 10.1109/MWC.001.2000267.

[2] R. Deng *et al.*, "Reconfigurable holographic surfaces for future wireless communications," *IEEE Wireless Communications*, vol. 28, no. 6, pp. 126–131, 2021.

[3] R. Deng, Y. Zhang, H. Zhang, B. Di, H. Zhang, and L. Song, "Reconfigurable Holographic Surface: A New Paradigm to Implement Holographic Radio," *IEEE Vehicular Technology Magazine*, vol. 18, no. 1, pp. 20–28, 2023, doi: 10.1109/MVT.2022.3233157.

[4] R. Deng *et al.*, "Reconfigurable Holographic Surfaces for Ultra-Massive MIMO in 6G: Practical Design, Optimization and Implementation," *IEEE Journal on Selected Areas in Communications*, 2023.

[5] G.-B. Wu *et al.*, "A universal metasurface antenna to manipulate all fundamental characteristics of electromagnetic waves," *Nature Communications*, vol. 14, no. 1, p. 5155, 2023.

[6] L. Zhang and T. J. Cui, "Space-time-coding digital metasurfaces: Principles and applications," *Research*, 2021.

[7] L. Zhang *et al.*, "A wireless communication scheme based on space- and frequency-division multiplexing using digital metasurfaces," *Nature Electronics*, vol. 4, no. 3, pp. 218–227, 2021, doi: 10.1038/s41928-021-00554-4.

[8] J. Park, J. H. Kang, S. J. Kim, X. Liu, and M. L. Brongersma, "Dynamic reflection phase and polarization control in metasurfaces," *Nano Letters*, vol. 17, no. 1, pp. 407–413, 2017, doi: 10.1021/acs.nanolett.6b04378.

[9] T. J. Cui, M. Q. Qi, X. Wan, J. Zhao, and Q. Cheng, "Coding metamaterials, digital metamaterials and programmable metamaterials," *Light: Science & Applications*, vol. 3, no. 10, pp. e218–e218, 2014, doi: 10.1038/lsa.2014.99.

[10] H. Yang *et al.*, "A programmable metasurface with dynamic polarization, scattering and focusing control," *Scientific reports*, vol. 6, no. 1, p. 35692, 2016.

[11] J. ur R. Kazim *et al.*, "In-Home Monitoring Using Wireless on the Walls for Future HealthCare: Real-World Demonstration," *Advanced Intelligent Systems*, p. 2300007, 2023.

[12] T. J. Cui, S. Liu, G. D. Bai, and Q. Ma, "Direct Transmission of Digital Message via Programmable Coding Metasurface," *Research*, vol. 2019, 2019, doi: 10.34133/2019/2584509.

[13] H. Li, M. Nerini, S. Shen, and B. Clerckx, "A Tutorial on Beyond-Diagonal Reconfigurable Intelligent Surfaces: Modeling, Architectures, System Design and Optimization, and Applications," *arXiv preprint arXiv:2505.16504*, 2025.

[14] Y. Zheng *et al.*, "Metasurface-Assisted Wireless Communication with Physical Level Information Encryption," *Advanced Science*, vol. 9, no. 34, p. 2204558, 2022.

[15] J. W. Xu, M. Wei, L. Zhang, V. Galdi, L. Li, and T. J. Cui, "Chaotic information metasurface for direct physical-layer secure communication," *Nature Communications*, vol. 16, no. 1, p. 5853, 2025.

[16] H. Q. Yang *et al.*, "Adaptively programmable metasurface for intelligent wireless communications in complex environments," *Nature Communications*, vol. 16, no. 1, p. 6070, 2025.

[17] J. W. Zhang *et al.*, "A smart millimeter-wave base station for 6G application based on programmable metasurface," *National Science Review*, vol. 12, no. 4, p. nwaf017, 2025.

[18] J. Y. Dai *et al.*, "Wireless communications through a simplified architecture based on time-domain digital coding metasurface," *Advanced materials technologies*, vol. 4, no. 7, p. 1900044, 2019.

[19] R. W. Shao *et al.*, "Dual-channel near-field holographic MIMO communications based on programmable digital coding metasurface and electromagnetic theory," *Nature Communications*, vol. 16, no. 1, p. 915, 2025.

[20] H. Xu, J. W. Wu, Z. X. Wang, R. W. Shao, H. Q. Yang, and T. J. Cui, "Two-dimensional and high-order directional information modulations for secure communications based on programmable metasurface," *Nature Communications*, vol. 15, no. 1, p. 6140, 2024.

[21] X. Y. Li *et al.*, "BioMeta: modular reprogrammable metasurface for noninvasive human respiration monitoring," *Nanophotonics*, vol. 14, no. 7, pp. 981–991, 2025.

[22] S. Hu *et al.*, "Electromagnetic metamaterial agent," *Light: Science & Applications*, vol. 14, no. 1, p. 12, 2025.

[23] Z. J. Qi *et al.*, "Over-The-Air Fast Diagnosis Strategy and Framework of Digital Coding Metasurfaces," *Advanced Materials Technologies*, p. e00892, 2025.

[24] S. R. Wang *et al.*, "Simplified radar architecture based on information metasurface," *Nature Communications*, vol. 16, no. 1, p. 6505, 2025.

[25] X. Q. Chen *et al.*, "Integrated sensing and communication based on space-time-coding metasurfaces," *Nature Communications*, vol. 16, no. 1, p. 1836, 2025.

[26] H. Taghvaee, A. Pitilakis, M. Khodadadi, O. Tsilipakos, A.-A. A. Boulogeorgos, and M. Khalily, "Holographic mmWave Metasurface Integrating THz Sensing for 6G Wireless Networks," *IEEE Wireless Communications*, vol. 32, no. 4, pp. 54–62, 2025.

[27] F. Verde, V. Galdi, L. Zhang, and T. J. Cui, "Integrating Sensing and Communications: Simultaneously Transmitting and Reflecting Digital Coding Metasurfaces: A successful convergence of physics and signal processsing [Special Issue on Signal Processing for the Integrated Sensing and Communications Re," *IEEE Signal Processing Magazine*, vol. 41, no. 5, pp. 56–70, 2024.

[28] B. Ji *et al.*, "Several key technologies for 6G: challenges and opportunities," *IEEE Communications Standards Magazine*, vol. 5, no. 2, pp. 44–51, 2021.

[29] M. Ahmed *et al.*, "A survey on STAR-RIS: Use cases, recent

advances, and future research challenges," *IEEE Internet of Things Journal*, vol. 10, no. 16, pp. 14689–14711, 2023.

[30] D. R. Smith, O. Yurduseven, L. P. Mancera, P. Bowen, and N. B. Kundtz, "Analysis of a waveguide-fed metasurface antenna," *Physical Review Applied*, vol. 8, no. 5, p. 54048, 2017.

[31] M. Boyarsky, T. Sleasman, M. F. Imani, J. N. Gollub, and D. R. Smith, "Electronically steered metasurface antenna," *Scientific reports*, vol. 11, no. 1, pp. 1–10, 2021.

[32] F. Yang *et al.*, "An End-Fed Programmable Metasurface Based on Substrate Integrated Waveguide for Novel Phased Array," *Advanced Engineering Materials*, vol. 25, no. 9, p. 2201251, 2023.

[33] R. Deng, B. Di, H. Zhang, Y. Tan, and L. Song, "Reconfigurable Holographic Surface-Enabled Multi-User Wireless Communications: Amplitude-Controlled Holographic Beamforming," *IEEE Transactions on Wireless Communications*, vol. 21, no. 8, pp. 6003–6017, 2022, doi: 10.1109/TWC.2022.3144992.

[34] B. Di, S. Zhang, Z. Han, and R. Zhang, "Holographic Beamforming with Leakage Power Constraints," *IEEE Wireless Communications*, vol. 32, no. 4, pp. 14–21, 2025.

[35] I. Yoo and D. R. Smith, "Dynamic Metasurface Antennas for Higher-Order MIMO Systems in Indoor Environments," *IEEE Wireless Communications Letters*, vol. 9, no. 7, pp. 1129–1132, 2020, doi: 10.1109/LWC.2020.2983374.

[36] N. Shlezinger, O. Dicker, Y. C. Eldar, I. Yoo, M. F. Imani, and D. R. Smith, "Dynamic Metasurface Antennas for Uplink Massive MIMO Systems," *IEEE Transactions on Communications*, vol. 67, no. 10, pp. 6829–6843, 2019, doi: 10.1109/TCOMM.2019.2927213.

[37] E. Björnson, F. Kara, N. Kolomvakis, A. Kosasih, P. Ramezani, and M. B. Salman, "Enabling 6G Performance in the Upper Mid-Band by Transitioning From Massive to Gigantic MIMO," *arXiv preprint arXiv:2407.05630*, 2024.

[38] H. Wang *et al.*, "Dynamic metasurface antennas for MIMO-OFDM receivers with bit-limited ADCs," *IEEE transactions on communications*, vol. 69, no. 4, pp. 2643–2659, 2020.

[39] R. Deng, B. Di, H. Zhang, Y. Tan, and L. Song, "Reconfigurable Holographic Surface: Holographic Beamforming for Metasurface-Aided Wireless Communications," *IEEE Transactions on Vehicular Technology*, vol. 70, no. 6, pp. 6255–6259, 2021, doi: 10.1109/TVT.2021.3079465.

[40] H. Zhang, H. Zhang, B. Di, and L. Song, "Holographic integrated sensing and communications: Principles, technology, and implementation," *IEEE Communications Magazine*, vol. 61, no. 5, pp. 83–89, 2023.

[41] W.-J. Lu *et al.*, "Fluid Antennas: Reshaping Intrinsic Properties for Flexible Radiation Characteristics in Intelligent Wireless Networks," *arXiv preprint arXiv:2501.02911*, 2025.

[42] W. K. New *et al.*, "A tutorial on fluid antenna system for 6G networks: Encompassing communication theory, optimization methods and hardware designs," *IEEE Communications Surveys & Tutorials*, 2024.

[43] R. J. Williams, P. Ram\'\irez-Espinosa, J. Yuan, and E. De Carvalho, "Electromagnetic based communication model for dynamic metasurface antennas," *IEEE Transactions on Wireless Communications*, vol. 21, no. 10, pp. 8616–8630, 2022.

[44] S. Li, F. Xu, X. Wan, T. J. Cui, and Y. Q. Jin, "Programmable Metasurface Based on Substrate-Integrated Waveguide for Compact Dynamic-Pattern Antenna," *IEEE Transactions on Antennas and Propagation*, vol. 69, no. 5, pp. 2958–2962, 2021, doi: 10.1109/TAP.2020.3023581.

[45] G. Lan *et al.*, "MetaSense: Boosting RF Sensing Accuracy Using Dynamic Metasurface Antenna," *IEEE Internet of Things Journal*, vol. 8, no. 18, pp. 14110–14126, 2021, doi: 10.1109/JIOT.2021.3070225.

[46] G. Lan, M. F. Imani, P. Del Hougne, W. Hu, D. R. Smith, and M. Gorlatova, "Wireless sensing using dynamic metasurface antennas: Challenges and opportunities," *IEEE Communications Magazine*, vol. 58, no. 6, pp. 66–71, 2020.

[47] S. X. Huang *et al.*, "Enabling Real-Time Near-Field Focusing Imaging with Space-Time-Coding Metasurface Antenna," *IEEE Transactions on Antennas and Propagation*, vol. 72, no. 12, pp. 9082–9094, 2024, doi: 10.1109/TAP.2024.3484665.

[48] B. Liu, K.-F. Tong, K.-K. Wong, C.-B. Chae, and H. Wong, "Be Water, My Antennas: Riding on Radio Wave Fluctuation in Nature for Spatial Multiplexing using Programmable Meta-Fluid Antenna," *arXiv preprint arXiv:2502.04693*, 2025.

[49] P.-Y. Wang, B. Sievert, A. Rennings, and D. Erni, "A Liquid Crystal-Based Dynamic Metasurface Antenna for Electronical Beam-Steering at 105 GHz," *IEEE Transactions on Antennas and Propagation*, 2024.

[50] A. Jabbar *et al.*, "60 GHz Programmable Dynamic Metasurface Antenna (DMA) for Next-Generation Communication, Sensing and Imaging Applications: From Concept to Prototype," *IEEE Open Journal of Antennas and Propagation*, 2024, [Online]. Available: doi: 10.1109/OJAP.2024.3386452

[51] S. Kutty and D. Sen, "Beamforming for millimeter wave communications: An inclusive survey," *IEEE communications surveys \& tutorials*, vol. 18, no. 2, pp. 949–973, 2015.

[52] S. Saponara, F. Giannetti, B. Neri, and G. Anastasi, "Exploiting mm-wave communications to boost the performance of industrial wireless networks," *IEEE Transactions on Industrial Informatics*, vol. 13, no. 3, pp. 1460–1470, 2017.

[53] J. G. Andrews, T. E. Humphreys, and T. Ji, "6 G Takes Shape," *IEEE BITS the Information Theory Magazine*, 2024.

[54] M. Luvisotto, Z. Pang, and D. Dzung, "Ultra High Performance Wireless Control for Critical Applications: Challenges and Directions," *IEEE Transactions on Industrial Informatics*, vol. 13, no. 3, pp. 1448–1459, 2017, doi: 10.1109/TII.2016.2617459.

[55] M. Luvisotto, Z. Pang, and D. Dzung, "High-performance wireless networks for industrial control applications: New targets and feasibility," *Proceedings of the IEEE*, vol. 107, no. 6, pp. 1074–1093, 2019.

[56] M. Luvisotto, Z. Pang, D. Dzung, M. Zhan, and X. Jiang, "Physical Layer Design of High-Performance Wireless Transmission for Critical Control Applications," *IEEE Transactions on Industrial Informatics*, vol. 13, no. 6, pp. 2844–2854, 2017, doi: 10.1109/TII.2017.2703116.

[57] M. Noor-A-Rahim *et al.*, "Toward Industry 5.0: Intelligent Reflecting Surface in Smart Manufacturing," *IEEE Communications Magazine*, vol. 60, no. 10, pp. 72–78, 2022.

[58] H. Ren, K. Wang, and C. Pan, "Intelligent Reflecting Surface-Aided URLLC in a Factory Automation Scenario," *IEEE Transactions on Communications*, vol. 70, no. 1, pp. 707–723, 2022, doi: 10.1109/TCOMM.2021.3125057.

[59] Z. Pang, M. Luvisotto, and D. Dzung, "Wireless high-performance communications: The challenges and opportunities of a new target," *IEEE Industrial Electronics Magazine*, vol. 11, no. 3, pp. 20–25, 2017.

[60] G. Lipworth *et al.*, "Metamaterial apertures for coherent computational imaging on the physical layer," *JOSA A*, vol. 30, no. 8, pp. 1603–1612, 2013.

[61] H. Xu, V. Kukshya, and T. S. Rappaport, "Spatial and temporal characteristics of 60-GHz indoor channels," *IEEE Journal on selected areas in communications*, vol. 20, no. 3, pp. 620–630, 2002.